\documentclass[journal=jacsat,manuscript=article]{achemso}
\usepackage[version=3]{mhchem}  

\usepackage{color}
\usepackage{amsmath}

\usepackage{soul}

\author{Tarak Karmakar$^{1,2,4}$}
\author{Michele Invernizzi$^{1,3,4}$}
\author{Valerio Rizzi$^{1,2,4}$}
\author{Michele Parrinello$^{1,2,4}$}
\affiliation
{1. Institute of Computational Sciences, Faculty of Informatics, Università della Svizzera italiana, Via Giuseppe Buffi 13, 6900 Lugano, Switzerland\\
2. Department of Chemistry and Applied Biosciences, ETH Zurich, 8092 Zurich, Switzerland\\
3. Department of Physics, ETH Zurich, 8092 Zurich, Switzerland\\
4. Italian Institute of Technology, Via Morego 30, 16163 Genova, Italy}
\email{michele.parrinello@phys.chem.ethz.ch}
\phone{+41 58 666 48 01}

\title[An \textsf{achemso} demo]
{Collective Variables for the Study of Crystallization}


\begin{document}

\newpage
\begin{abstract}
The phenomenon of solidification of a substance from its liquid phase is of the greatest practical and theoretical importance, and atomistic simulations can provide precious information towards its understanding and control. Unfortunately, the time scale for crystallization is much larger than what can be explored in standard simulations. Enhanced sampling methods can overcome this time scale hurdle. Here we employ the on-the-fly probability enhanced sampling method that is a recent evolution of metadynamics. This method, like many others, relies on the definition of appropriate collective variables able to capture the slow degrees of freedom. To this effect we introduce collective coordinates of general applicability to crystallization simulations. They are based on the peaks of the three-dimensional structure factor that are combined non-linearly via the Deep Linear Discriminant Analysis machine learning method. We apply the method to the study of crystallization of a multicomponent system, Sodium Chloride and a molecular system, Carbon Dioxide.
\end{abstract}

\newpage
\section{1. Introduction}
Understanding at the microscopic level the process of crystallization is of great importance in many areas of physics, chemistry, and biology.\cite{hulliger1994chemistry,schuler1999bacterial, sloan2003fundamental,doherty2006crystal,erdemir2007polymorph} From a more general point of view, crystallization provides a spectacular example of symmetry-breaking first-order phase transition. In principle, atomistic molecular dynamics simulations could offer invaluable insight, given the fact that experimental studies are rather challenging. This possibility did not escape the attention of the early simulators, and already in 1976, Rahman studied the homogeneous crystallization of a Lennard-Jones liquid.\cite{mandell1976crystal} However, since crystallization takes place in a macroscopic time scale, Rahman was forced to apply deep and unrealistically fast temperature quenchings in order to induce crystallization.\cite{mandell1976crystal,mandell1977crystal} It became then clear that in order to study crystallization, enhanced sampling methods are mandatory.\cite{auer2001prediction,ten1997enhancement,tribello2009molecular,giberti2013transient,giberti2015insight,salvalaglio2013controlling,giberti2015metadynamics} Such methods allow overcoming the large free energy barriers involved in the crystallization process, thus circumventing the presence of kinetic bottlenecks. 

The overwhelming majority of such studies have used methods that directly or indirectly require the definition of a small set of collective variables (CVs).\cite{giberti2013transient,giberti2015insight,salvalaglio2013controlling,piaggi2017enhancing,piaggi2018predicting,gobbo2018nucleation,niu2018molecular,bonati2018silicon,piaggi2020phase,niu2020ab} The CVs are meant to encode the slow modes of the system such that, by applying a bias potential that is a function of the CVs, it is possible to remove the kinetic bottlenecks. Designing such CVs is often a non-trivial task and a successful choice of CVs demands physical intuition and an in-depth understanding of the system studied. Practical considerations suggest limiting the number of CVs used, and much ingenuity has been invested into condensing several physical parameters into a small number of CVs.\cite{santiso2011general,giberti2015insight,zhang2019improving}

The CVs can be categorized into three classes. In the first class, one could include generic CVs such as enthalpy and an entropy surrogate whose aim is to drive crystallization without prior knowledge of the crystal structure.\cite{piaggi2017enhancing,piaggi2018predicting} These CVs are useful when exploring polymorphism, but they are not very efficient when it comes to calculating free energies. If the crystal structure is known, a second class of CVs has been designed that describes the local crystalline environment.\cite{rein1996numerical,lechner2008accurate,santiso2011general,giberti2015insight,piaggi2019calculation,piaggi2020phase} These CVs have been very useful but it is difficult to extend their application to multicomponent systems, and if not properly chosen, they can drive the system towards unwanted defective structures. A third class of CVs, that also involves knowledge of the final crystal structure, is obtained by analyzing the atomic density in the reciprocal Fourier space and using as CVs the Bragg peak intensities that are the experimental signature of crystallinity. Guided by this consideration, we have recently used the peaks in the shperically-averaged structure factor as possible CVs for crystallization.\cite{niu2018molecular,bonati2018silicon,Invernizzi2019making} When applying this idea in the practice, one usually needs to reduce the number of peaks considered e.g., by linearly combining different peaks with a lengthy trial and error procedure, as done in Refs. \citenum{zhang2019improving} and \citenum{niu2019temperature}.

In this paper, instead of using the peaks of the spherically-averaged structure factor following the Debye scattering equation as in Ref.~\citenum{niu2018molecular}, we propose to use the full three-dimensional structure factor (SF).
This has two important advantages.
On one hand, it provides a more accurate description by resolving each scattering vector into its three-dimensional components. This avoids the degeneracy intrinsic to the Debye peaks, which can lead to mislabeling defected structures. On the other hand, the SF is also computationally more efficient, since it scales linearly with the number of atoms. However, an inevitable hurdle of this choice is that the 3D SF presents a number of crystal peaks much greater than its spherically-averaged counterpart. Thus it becomes crucial to avoid the lengthy procedure of Ref.~\citenum{niu2019temperature} and replace it by a modern machine learning (ML) technique, in which several descriptors are non-linearly combined to produce efficient CVs.\citep{rogal2019neural,bonati2020data} 
To this effect, we use the Deep Linear Discrimination Analysis (Deep-LDA) method\citep{bonati2020data} and obtain a Deep-LDA CV that we then use in the On-the-fly Probability Enhanced Sampling (OPES) method\citep{invernizzi2020rethinking,invernizzi2020unified} to simulate the crystallization of multicomponent and molecular systems. As prototypical examples, we have applied with success our approach to study the crystallization of NaCl and CO$_2$ from their liquid phase. We have chosen these systems since they present a number of challenges as discussed in the literature.\citep{giberti2013transient,valeriani2005rate,gimondi2017co2,gimondi2018co}

\newpage
\subsection{2. OPES}
In order to accelerate sampling, we use the recently developed OPES method.\citep{invernizzi2020rethinking,invernizzi2020unified} 
In OPES, like in metadynamics\citep{laio2002escaping,barducci2008well,barducci2011metadynamics,valsson2016enhancing}, the bias is constructed on-the-fly with the help of periodically added Gaussians $G(s,s_n)$ that are functions of the chosen CVs $s$ and are centered at the CVs current $s_n$ value, where $n$ labels the $n^{th}$ deposition. 
In metadynamics, the Gaussians are directly added to the bias potential $V(s)$. In OPES instead, the added Gaussians modify the equilibrium probability distribution $P(s)$ which in turn is related to the bias by:
\begin{equation}
    V(s) = - \frac{1}{\beta}\log\frac{p^{\text{tg}}(s)}{P(s)}
\end{equation}
in which $\beta^{-1} = {k_BT}$ is the inverse temperature, and $p^{\text{tg}}(s)$ is the target distribution we want the OPES simulation to converge to. In the present application, we will aim at obtaining the so-called well-tempered distribution $p^{\text{tg}}(s) \propto [P(s)]^{\frac{1}{\gamma}}$, where $\gamma > 1$ is called bias factor. The merits of this choice have been amply discussed in the metadynamics literature.\cite{barducci2008well,barducci2010linking,dama2014well}

In order to compute $P(s)$, one sets up an iterative scheme in which at the $n^{th}$ step the probability distribution is estimated as:
\begin{equation}
    P_n(s) = \frac{\sum_k^n w_k G(s,s_k)}{\sum_k^n w_k}
\end{equation}
and the weights $w_k = e^{\beta V_{k-1}(s_k)}$ are given by the bias potential previously deposited.

The relation between the $n^{th}$ estimate of $P_n(s)$ and that of $V_n(s)$ is given by
\begin{equation}
    V_n(s) = (1-1/\gamma)\frac{1}{\beta}\log\left ( \frac{P_n(s)}{Z_n} + \epsilon \right )
\end{equation}
where $Z_n$ is a normalization factor, and $\epsilon$ is a regularization parameter that sets a maximum limit to the bias potential, in order to avoid exploring too high free energy regions.

The $n^{th}$ free energy estimate then becomes
\begin{equation}
    F_n(s) = -\frac{1}{\beta}\log {P_n(s)}
\end{equation}
Similarly, any other ensemble average of an observable $O=O(\mathbf{R})$ that is function of the coordinates $\mathbf{R}$, can be estimated via reweighting
\begin{equation}
    \langle O \rangle= \frac{\langle O\, e^{\beta V}\rangle_V}{\langle e^{\beta V}\rangle_V}
\end{equation}
where $\langle \cdot \rangle_V$ denotes the ensemble average over the biased simulation.

The discussion presented here is highly schematic and misses many of the important technical details that make OPES efficient. These can be found in Refs \citenum{invernizzi2020rethinking} and \citenum{invernizzi2020unified}.

\section{3. LDA and Deep-LDA}
In order to design efficient CVs, we follow the approach of Ref. \citenum{bonati2020data} that relies on the linear discriminant analysis (LDA) introduced a long time ago by Fisher\cite{welling2005fisher}. We specialize our discussion to the case in which only two metastable states are present since here we are only interested in studying transitions between solid (S) and liquid (L) phases. For a treatment of the more general case in which several metastable states are present the interested reader is referred to the literature.\cite{bonati2020data}

The transition between solid and liquid is a rare event occurring on a time scale that far exceeds that of standard simulations. Thus, if we perform two distinct simulations one in the solid and the other in the liquid, no transition will be observed, and the configurations thus generated will be separated in the atomic coordinate multidimensional space. We want to project these data into a low dimensional set of CV that is still able to discriminate between the two states. As stated earlier, here we shall apply a method derived from LDA.

In LDA, one starts by selecting an exhaustive set of descriptors that are $N_d$ functions of the atomic coordinates $\boldsymbol{R}$. For notational convenience we arrange the $d_i(\boldsymbol{R})$ to form a vector $\boldsymbol{d(R)}$, and in the following, we shall not make explicit the $\boldsymbol{R}$ dependence of $\boldsymbol{d(R)}$. The descriptors could in theory be all the $\boldsymbol{R}$, but it is better to preprocess the $\boldsymbol{R}$ such that the result remains physically transparent, and the natural symmetries of the system are respected. For instance, one could use as descriptors distances or angles that have the property of being translationally and rotationally invariant. In addition, for the purpose of generating useful CVs, the descriptors should be able to correctly distinguish between the various metastable and transition states. This requirement is not necessarily satisfied by a CV that is generated in a discrimination procedure. However, experience has shown that this is to a good extent the case. 

The object of LDA is to find a linear combination of descriptors $s=\mathbf{w^Td}$ such that when the data are projected on $s$ the variance of S and L data is large while the total spread is low. In order to obtain the optimal $\mathbf{w}$, one first runs two distinct simulations, one in the solid and the other in the liquid. From the data thus collected one computes the averages and covariances of the descriptors in each state and arrange them to form the vectors $\boldsymbol{d_S}$ and $\boldsymbol{d_L}$ and matrices $\boldsymbol{\Sigma_S}$ and $\boldsymbol{\Sigma_L}$, respectively. These quantities are combined to form two $N_d \times N_d$ covariance matrices, the within $\boldsymbol{S}_w = \frac{1}{2}\left(\boldsymbol{\Sigma_S} + \boldsymbol{\Sigma_L} \right)$ and the between $\boldsymbol{S}_b = \boldsymbol{(d_S-d_L)(d_S-d_L)^T}$ one. The former measures the total spread of the projected distribution while the latter informs on its total variance. 

The condition of high separation in the $s$ projection between data pertaining to the different states and low variance within each state is obtained by maximizing the Fisher ratio:
\begin{align}
    \frac{\boldsymbol{\mathbf{w} S_b \mathbf{w}^T}} {\boldsymbol{\mathbf{w} S_w \mathbf{w}^T}}
    \label{eq:Fish}
\end{align}
with respect to $\mathbf{w}$. The value of $\mathbf{w}^*$ that maximizes the Fisher ratio can be obtained as the eigenvector of the highest eigenvalue of the generalized eigenvalue equation.
\begin{align}
\boldsymbol{S_b\mathbf{w_i}} = \nu_i \boldsymbol{S_w\mathbf{w_i}}
    \label{eq:fisheq}
\end{align}

The sought-after CV is then
\begin{align}
    s=\boldsymbol{\mathbf{w}^{*T}d}
    \label{eq:ldacv}
\end{align}

In Deep-LDA, the $\boldsymbol{d}$ is first inputted into a feed-forward deep neuronal network (NN), whose output $\boldsymbol{h}$ has dimension $N_h$. One then starts with a guess for $\boldsymbol{h}$, calculates $\boldsymbol{S_b}$ and $\boldsymbol{S_w}$ that, in the $\boldsymbol{h}$ basis, are now $N_h \times N_h$ matrices. The NN weights are optimized using as loss function the eigenvector with lowest eigenvalue ($\nu$) of the Fisher's generalized equation Eq. \ref{eq:fisheq} that is now a $N_h \times N_h$ equation, this leads to a new $\boldsymbol{h}$ and to a new optimization procedure. The cycle is repeated until convergence leading to the Deep-LDA CV, 
\begin{align}
    s=\boldsymbol{\mathbf{w}^Th}
    \label{eq:deepcv}
\end{align}
This non-linear procedure has proven to lead to more efficient CVs than the one based on LDA even when using its harmonic version HLDA.\cite{mendels2018collective}

\newpage
\section{4. Collective Variable}
As discussed in the previous section we want to combine a set of descriptors in a non-linear way so as to obtain efficient CVs for the study of crystallization. Typically, crystal order parameters are defined locally, by considering the environment of each atom up to a given cutoff radius.
We introduce instead a descriptor capable to take into account for the long-range order of the crystal.
Following Ref.~\citenum{niu2018molecular} we propose to use the structure factor as descriptor, but contrary to Ref.~\citenum{niu2018molecular}, here we will not use the isotropic Debye formula, but rather the full three-dimensional structure factor, as defined for a monoatomic substance
\begin{equation}\label{eq:sk}
    S({\boldsymbol{k}}) = \frac{1}{N}\left\lvert \sum_{i=1}^{N} e^{-i\boldsymbol{k}\cdot \boldsymbol{R_i}}\right\rvert^2
\end{equation}
where $\mathbf{k}$ is the three-dimensional scattering vector.
For simplicity, we consider a periodically repeating cubic simulation box of edge $L$, so $\mathbf{k}$ can be written as
\begin{equation}
    \boldsymbol{k} = \frac{2\pi}{L}(l,m,n)
\end{equation}
with $l,m,n$ integer indexes. In the case of a crystal structure that is commensurate to the MD cell, only a subset of $l,m,n$ will correspond to a Bragg peak. Our choice is to set ourselves in this condition and choose also the number of atoms such that perfect commensuration is possible. Out of simplicity, we are considering here only the case of a cubic box, but extension to more general type of periodic boundary conditions is straightforward.

The advantage of Eq.~\ref{eq:sk} is that, contrary to the Debye formula, it scales linearly with the number of atoms $N$ without the need to introduce a cutoff radius.
The main drawback is that the number of valid scattering vectors $\mathbf{k}$ grows roughly as $N^3$. 
This however, does not constitute a problem, because for any given crystal structure $S(\mathbf{k})$ will present only few high intensity peaks at specific values of $\mathbf{k}$ while all the others can be safely ignored. 
Notice that the peaks of the three-dimensional SF should not be confused with the peaks of the spherically-averaged SF, $S(|\mathbf{k}|)$, that is the one typically measured in powder X-ray diffraction experiments.

\begin{figure}[!hb]
\includegraphics[width=0.8\textwidth]{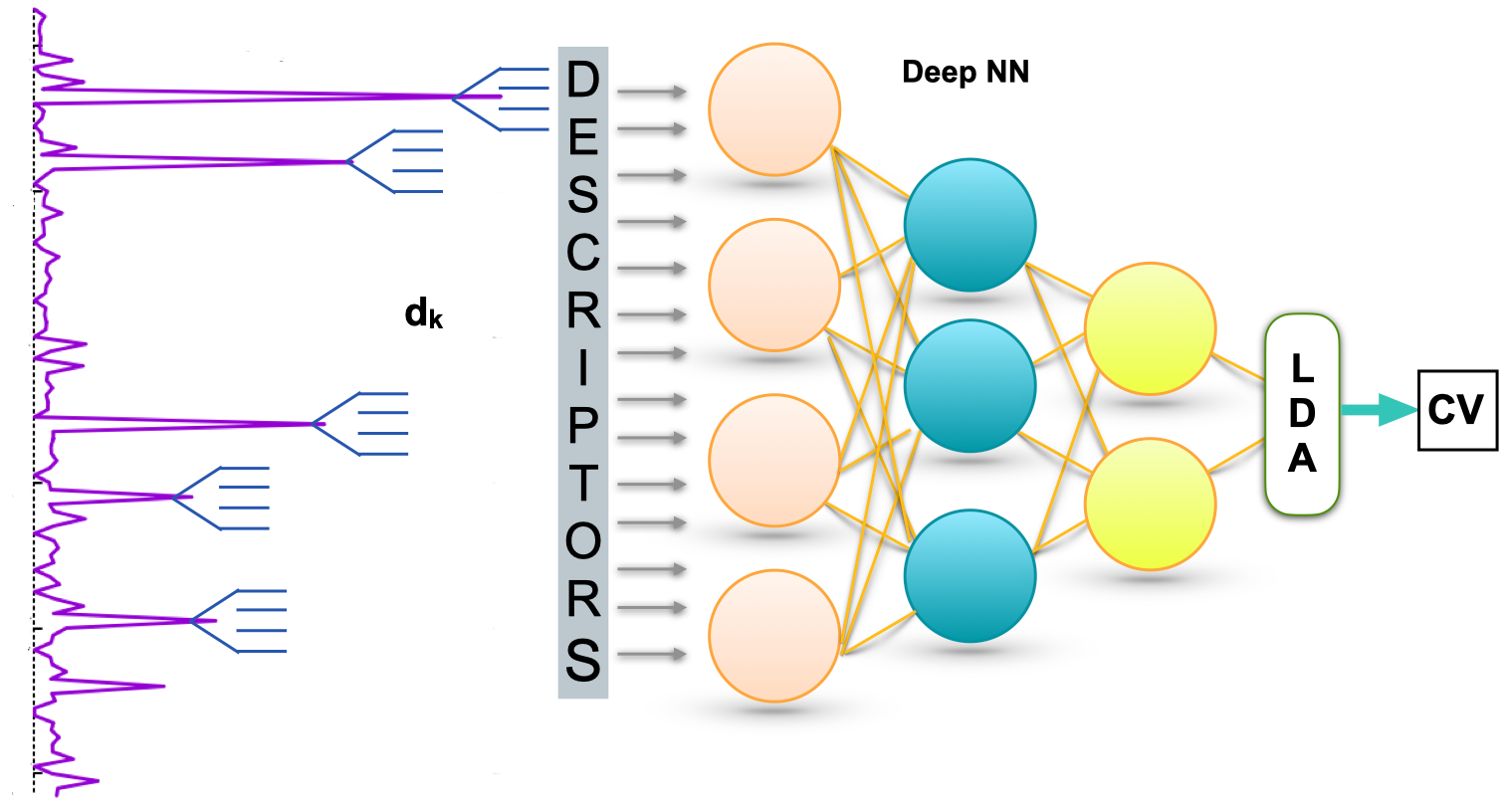}
	\centering
	\caption{Schematic of the Deep-LDA protocol, left: A representative spherically-averaged structure factor $S(|\boldsymbol{k}|)$ profile. Each of the shown peaks is constituted of several SF peaks, corresponding to scattering vectors $\mathbf{k}$ that have same modulus, but different components.	The corresponding descriptors ($\boldsymbol{d_k}$) are provided as input for the Deep-LDA network.} 
	\label{fig:scheme}
\end{figure}

In order to define the SF descriptors for a given system, we proceed as follows. First we run a short unbiased simulation of the perfect crystal and calculate the average $S(\mathbf{k})$.
We then consider as descriptors only those SF peaks that are greater than a threshold value $\varepsilon_I$ and whose scattering vector is smaller in modulus than a preassigned cutoff $|\mathbf{k}|<k_c$ where information on the X-ray crystal structure are damped by the Debye-Waller factor.
This typically still leaves us with a sizable number of descriptors, thus they cannot be used directly as CVs, but they can be combined via the Deep-LDA machine learning algorithm, in order to obtain a single CV, as shown schematically in Fig.~\ref{fig:scheme}.

For practical reasons, before feeding them to the neural network, we take the square root of the structure factor peaks, thus the descriptors are the following:
\begin{equation}
    d_{\mathbf{k}}=\sqrt{S(\mathbf{k})}=
    \frac{1}{\sqrt{N}} \left\lvert \sum_{i=1}^{N} e^{-i\boldsymbol{k\cdot R_i}}\right\rvert
    \label{eq:dk}
\end{equation}

An important characteristic of the proposed descriptors is that by selecting only the $\mathbf{k}$ vectors corresponding to a given crystal configuration, we obtain descriptors that are not rotationally invariant, but that favour the formation of a crystal perfectly aligned with the simulation box. This, as proposed in Ref.~\citenum{piaggi2019calculation}, is essential to avoid secondary defected states and greatly simplify and speedup the calculation.

\section{5. Simulation details}
\noindent
The crystal structures of NaCl and CO$_2$ were taken from the crystallographic database. The unit cells were replicated to generate the supercells using the AVOGADRO software.\cite{hanwell2012avogadro} In the subsequent step, the supercells were minimized and thermally equilibrated in constant temperature simulation. The corresponding liquid phases were obtained by melting the crystal at a temperature higher than their respective melting temperatures and afterward cooling down to the desired temperature. 

The systems were further equilibrated by carrying out short simulations in the isothermal-isobaric ($NPT$) ensemble. The temperature and pressure of the system were controlled by the stochastic velocity rescaling thermostat\cite{bussi2007canonical} and the Berendsen barostat,\cite{berendsen1984} respectively. For the production $NPT$ simulations, we retain the same thermostat but switch to an isotropic Parrinello-Rahman barostat.\cite{parrinello1981polymorphic} A time step of 2 fs is used in all production simulations. The coupling constants for the thermostat and barostat were 0.5 ps and 1 ps, respectively. A cut-off of 0.7 and 0.85 was used for both the van der Waals and the short-range Coulomb interactions for NaCl and CO$_2$ systems, respectively. Particle Mesh Ewald was utilized to treat the long-range electrostatic interactions. The LINCS algorithm was used to constraint the bonds. The Joung-Cheatham\cite{joung2008determination} and EPM2-flexible potential model\cite{epm2,zhang2005co2} parameters were used to simulate NaCl and CO$_2$, respectively.

The OPES bias uses adaptive kernel bandwidth and is updated every 1000 time steps in the case of NaCl and every 500 for CO$_2$, while the barrier parameter is set to 600 and 200 kJ/mol respectively.
All simulations were carried out in GROMACS-2019.4 software\cite{abraham2015gromacs} patched with a private version of the PLUMED2.7-dev plugin.\cite{tribello2014plumed} The in-house built codes and input files that were used in this work are available at PLUMED-NEST (...), a public repository of the PLUMED consortium.\cite{bonomi2019promoting} The figures were prepared using the Visual Molecular Dynamics (VMD) software.\cite{humphrey1996vmd}

\section{6. Results and discussion}
\subsection{6.1. Sodium Chloride}
While much attention has been paid to the nucleation of monatomic systems, multicomponent systems have been much less studied. As a representative example, we discuss here the case of NaCl. 

\begin{figure}[!ht]
\includegraphics[width=1.\textwidth]{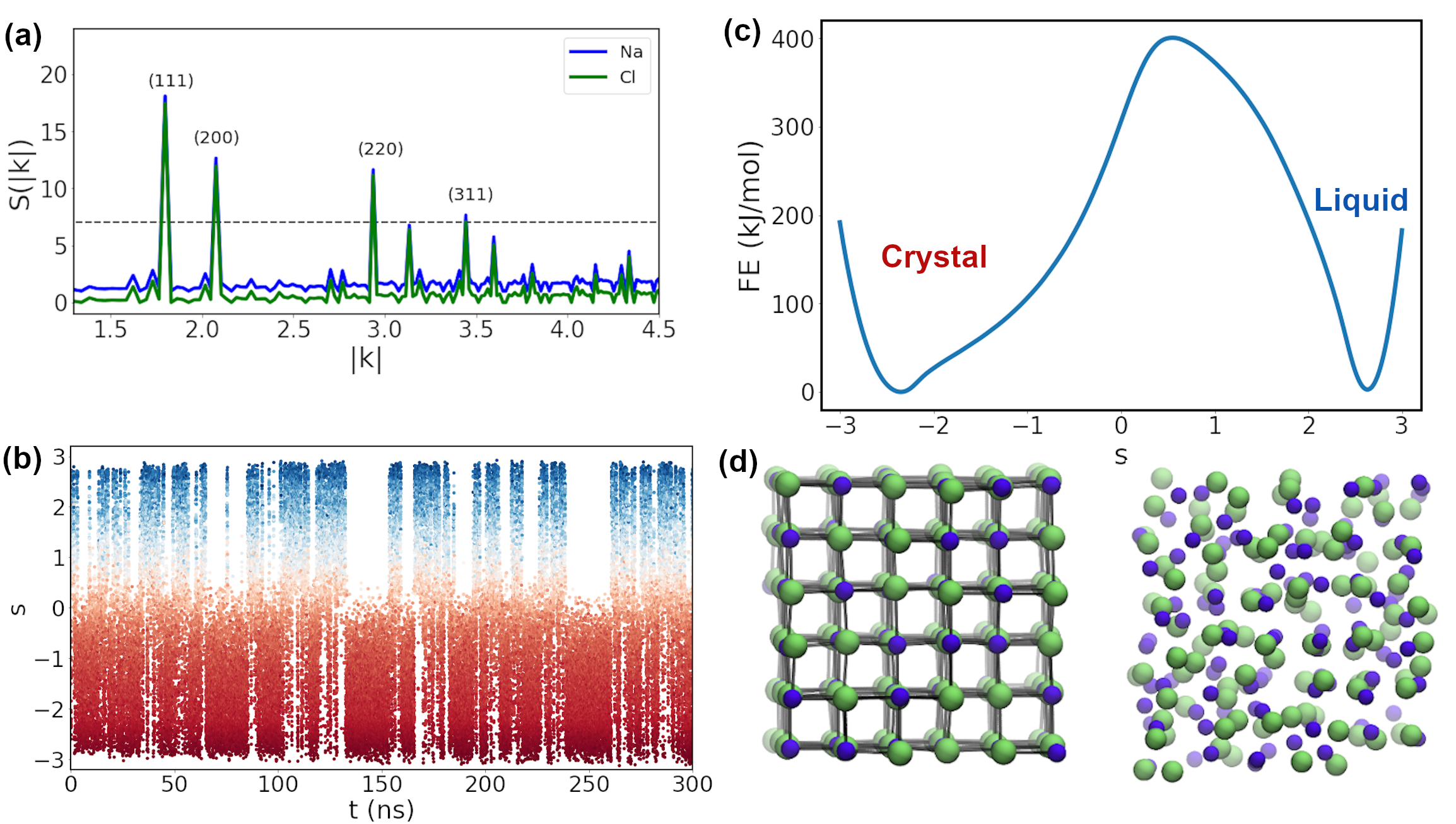}
	\centering
	\caption{(a) The $S(|\mathbf{k}|)$ profiles for Na$^+$ and Cl$^-$ ions. The two profiles are identical, so for visibility the Na$^+$ profile is shifted upward. (b) The time series of NaCl CV $s$ obtained from the OPES simulation. The color is based on the first descriptor that corresponds to the Miller index (111) used in the Deep-LDA CV, red: crystal and blue: liquid. and (c) The free energy profile as a function of $s$ for the crystallization process at 1250 K, two representative configurations, (d) rock-salt crystal and the liquid configurations are shown.}
	\label{fig:nacl}
\end{figure}

Following the scheme described earlier we perform an equilibrium simulation of NaCl in its rock salt structure at T=1250 K. For graphical convenience, we report in Fig. \ref{fig:nacl}(a) the spherically-averaged SF profile, $S(|\boldsymbol{k}|)$, as obtained from this simulation.
. 
We use the cutoff values (See section Collective variable) $k_c$ = 3.4 and $\varepsilon_I$ = 7 to select the $\boldsymbol{k}$ whose intensity will form our set of descriptors. The chosen $\boldsymbol{k}$'s correspond to the Miller indices (111), (200), (220) and (311). Considering that we introduce for each $\boldsymbol{k}$ orientation a different descriptor, this leads to 15 independent $d_\mathbf{k}$ (note that $\boldsymbol{k}$ and $-\boldsymbol{k}$ are equivalent). After having performed an unbiased simulations in the solid (S) and liquid (L), we can compute the averages and fluctuations of the $d_\mathbf{k}$’s in the S and L states and build from those the Deep-LDA CV as prescribed in Sec. 3. 

The fact that NaCl is a two-component system requires extending our scheme to a multi-component system. In a strong ionic system like NaCl in which establishing local neutrality is paramount, the two components Na$^+$ and Cl$^-$ will have a similar behavior. Thus, for each ionic species we determine its CV $s_{Na}$ and $s_{Cl}$ as described previously. Finally, the CV used in the OPES simulation is the linear combination, $s = 0.5*(s_{Na} + s_{Cl})$. We perform an OPES simulation with this CV at T 1250 K and P 1 bar.

A large number of reversible transitions between the solid and the liquid phase manifests the effectiveness of the CV, Fig. \ref{fig:nacl}(b), and allows us to calculate the converged free energy profile for the crystallization process. The free energy profile is shown in Fig. \ref{fig:nacl}(c), and the free energy difference ($\Delta F$) between the two states converges to $\sim$10 kJ/mol in $\sim$ 160 ns simulation time (see Fig. S2). At T = 1250 K, both the crystal and liquid basins have almost comparable free energies with a slightly deeper minimum for the crystal basin. This agrees with the fact that the chosen temperature, T = 1250 K is a little below the melting temperature of the the Joung-Cheatham potential model of NaCl which was found to be 1285 $\pm$ 5 K.\cite{joung2008determination,aragones2012calculation}

\subsection{6.2. Carbon dioxide}
\begin{figure}[!ht]
\includegraphics[width=1.0\textwidth]{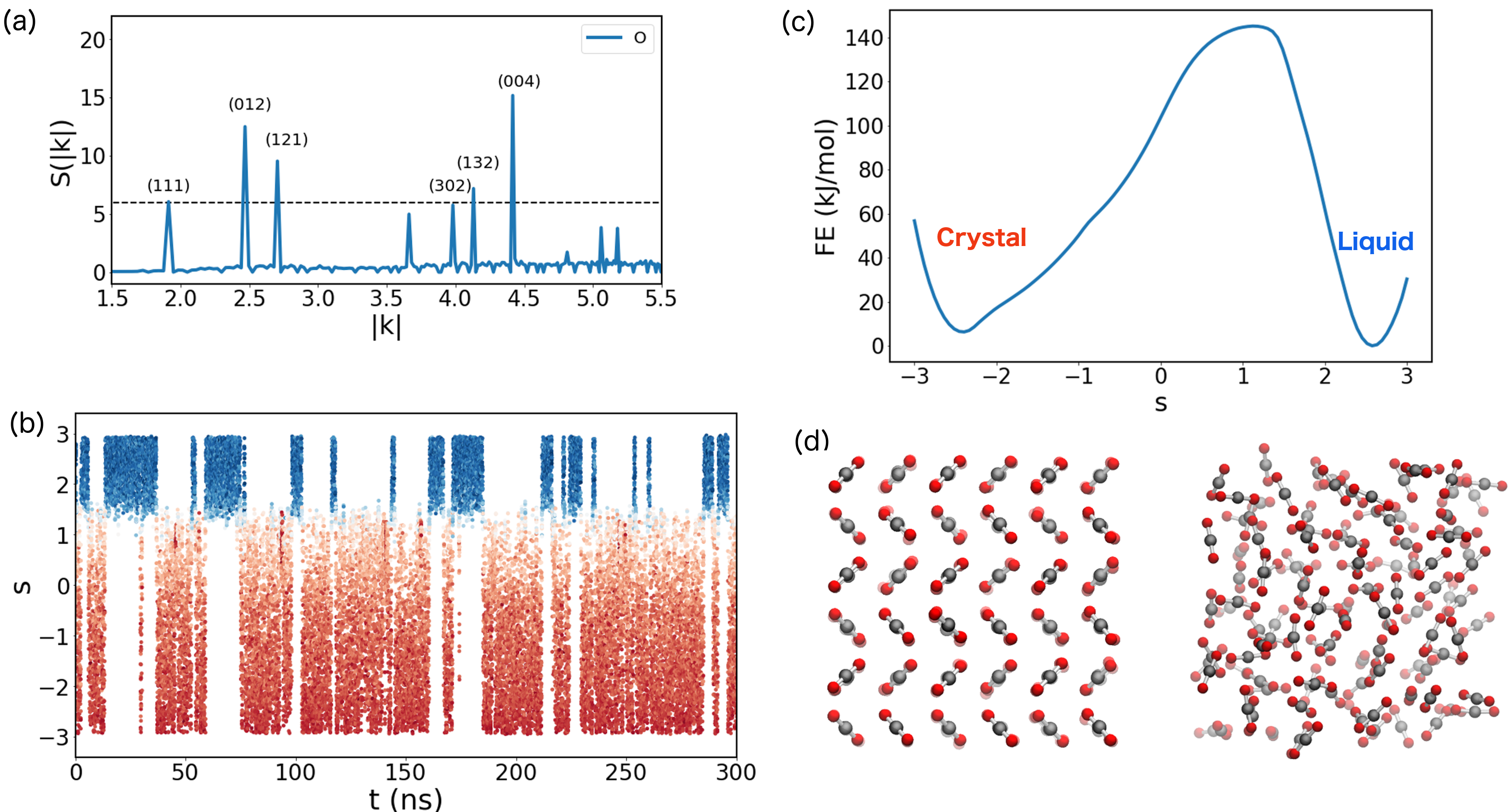}
	\centering
	\caption{(a) The $S(|\mathbf{k}|)$ profile of CO$_2$ crystal. (b) Time evolution of the CO$_2$ CV $s$ obtained from the OPES simulation. The color is based on the value of the first descriptor that corresponds to the Miller index (111) used in the Deep-LDA CV, red: crystal and blue: liquid. (c) The free energy profile for the crystallization process at T = 400 K and P = 2 GPa, and representative (d) crystal and liquid configurations are shown.} 
	\label{fig:co2}
\end{figure}

We extend our approach to study the crystallization of molecular systems and choose for this study, carbon dioxide. This seemingly simple system has a very rich phase diagram consisting of gas, compressed fluid, and several polymorphic crystal phases.\cite{datchi2009structure,santoro2006high,yoo2013physical,giordano2007molecular,iota2001phase,bonev2003high,etters1989static} At ambient conditions, CO$_2$ exists as a gas, and when cooled and compressed, it transforms into a dense liquid which can further be converted to a crystal, popularly known as dry ice.  In this work, we restrict ourselves to studying the phase transition from the liquid to its crystal polymorph-I (Fig. \ref{fig:co2}). We choose to simulate a 3$\times$3$\times$3 crystal super-cell containing 108 CO$_2$ molecules at T = 400 K and P = 2 GPa. These T and P values are close to the liquid-solid phase coexistence point.

In the CO$_2$ structure, it suffices to specify the oxygen positions to determine the full crystal structure. For this reason we build the CO$_2$ CV from the oxygen partial densities. Thus we calculate the $S_{OO}(\boldsymbol{k}$) partial structure factor with respect to the oxygen atoms (Fig. \ref{fig:co2}). Here we choose the peaks having $\varepsilon_I > 6$ with Miller indices (111), (012), (121), (302), (132), and (004) (Fig. \ref{fig:co2}(a)) are fed to the Deep-LDA network to obtain the CV. Subsequently, we have carried out an OPES simulation with this CV at T = 400 K and P = 2 GPa.

The system visits the S and L states several times, as can be seen in Fig. \ref{fig:co2}(b) and $\Delta F$ is converged within 300 ns simulation time (Fig. S4). The free energy profile is shown in Fig. \ref{fig:co2}(c). A nucleation barrier of $\sim$140 kJ/mol and a free energy difference of $\sim$5 kJ/mol between the two states is obtained. In agreement with the experiments, at the thermodynamic conditions studied here, the liquid state is slightly more stable than the crystal.

\section{7. Conclusions}
In this paper, we introduce the use of the structure factor peaks as descriptors for enhanced sampling simulations of crystallization events.
Contrary to the CVs previously proposed in Ref.~\citenum{niu2018molecular}, based on the Debye formula for spherically-averaged SF, our descriptors are non-degenerate and scale more efficiently with the system size. They also induce crystallization only along the box directions, similarly to the CVs proposed in Ref.~\citenum{piaggi2019calculation}.
These SF descriptors, combined with the Deep-LDA ML method, provide a systematic and efficient way for obtaining low dimensional CVs for a vast range of crystallization problems.

The effectiveness of our approach in simulation of crystallization is manifested by two representative examples, NaCl and CO$_2$. In each case, the Deep-LDA CV is found effective in reversibly transforming the system from its liquid to the crystal state in relatively short simulation time. This allows us to calculate the free energy of the liquid and crystal phase at the chosen temperature and pressure. Our method opens up new avenues to efficiently study the liquid-solid phase transition and estimate the stability of the crystal phase relative to its molten state. We believe that this strategy can be applied to study other phase transitions such solid-solid, amorphous-solid, and liquid-liquid transitions.

\begin{acknowledgement}
The research was supported by the NCCR MARVEL (D\&D1) funded by Swiss National Science Foundation and the European Union Grant No. ERC-2014-AdG-670227/VARMET. We thank CSCS, the Swiss National Supercomputing Centre for providing the computational resources. 

\end{acknowledgement}

\begin{suppinfo}
The Supporting Information (SI) contains free energy difference plots (Fig. S1 and S2) obtained from OPES simulations of NaCl and CO$_2$ systems.
\end{suppinfo}

\section{Code and data Availability}
All input files, codes, scripts, and instructions to reproduce the results presented in this manuscript can be found at the
PLUMED-NEST repository as plumID:XXX. Additionally, all simulations data are available from the authors upon request.

\newpage
\bibliography{main}

\end{document}


\newpage
\section{A. NaCl}
\begin{figure}[!h]
\includegraphics[width=.8\textwidth]{Figures/DF_nacl.png}
	\centering
	\caption{$\Delta F$ is plotted as a function of OPES simulation time. The FE converges after $\sim$150 ns.}
	\label{fig:nacl}
\end{figure}

\newpage
\section{A. CO$_2$}

\begin{figure}[!h]
\includegraphics[width=.8\textwidth]{Figures/DF_co2.png}
	\centering
	\caption{$\Delta F$ is plotted as a function of OPES simulation time. The FE converges after $\sim$120 ns.}
	\label{fig:nacl}
\end{figure} 

\newpage